\documentclass[preprint,showpacs,preprintnumbers,amsmath,amssymb]{revtex4}

\usepackage{graphicx}% Include figure files
\usepackage{dcolumn}% Align table columns on decimal point
\usepackage{bm}% bold math

\def\<#1|{\langle#1|}
\def\|#1>{|#1\rangle}

\def\T2{\alpha_{L/2}}

\def\bk<#1|#2>{\left\langle\vphantom{#1|#2}#1\right|%
\left.\vphantom{#1|#2}#2\right\rangle}
\def\kb|#1><#2|{\left|\vphantom{#1|#2}#1\right\rangle%
\left\langle\vphantom{#1|#2}#2\right|}
\def\aux|#1><#2|{\left|\vphantom{#1|#2}#1\right\rangle_{a}%
\left\langle\vphantom{#1|#2}#2\right|}
\def\matrix[#1][#2]{\left[
  \begin{array}{#1}
    #2
  \end{array}  \right]}

\def\a{\hat{\op{a}}}
\def\b{\hat{\op{b}}}
\def\c{\hat{\op{c}}}

\def\e{\hat{\op{e}}}
\def\f{\hat{\op{f}}}
\def\g{\hat{\op{g}}}
\def\h{\hat{\op{h}}}
\def\i{\hat{\op{i}}}

\newcommand{\op}[1]{{\mathbf #1}}

\def\sigz{\op{\sigma_z}}
\def\sigm{\op{\sigma_-}}
\def\sigp{\op{\sigma_+}}
\def\aout{\a_{out}}
\def\cout{\c_{out}}
\def\eout{\e_{out}}
\def\ain{\a_{in}}
\def\cin{\c_{in}}
\def\ein{\e_{in}}
\def\aeven{\a_{even}}
\def\aodd{\a_{odd}}
\def\domega{\Delta\omega}

\def\aoutdag{\a_{out}^{\dagger}}
\def\coutdag{\c_{out}^{\dagger}}

\def\aindag{\a_{in}^{\dagger}}

\def\Dalpha[#1]{D_\alpha\left(#1\right)}

\begin{document}

\preprint{APS/123-QED}

\title{Dipole Induced Transparency in drop-filter cavity-waveguide systems}% Force line breaks with \\

\author{Edo Waks}
\affiliation{
E.L. Ginzton Labs\\
Stanford University, Stanford, CA, 94305
}%

\author{Jelena Vuckovic}
\affiliation{
E.L. Ginzton Labs\\
Stanford University, Stanford, CA, 94305
}%

\date{\today}% It is always \today, today,

\begin{abstract}
We show that a waveguide that is normally opaque due to
interaction with a drop-filter cavity can be made transparent when
the drop filter is also coupled to a dipole.  A transparency
condition is derived between the cavity lifetime and vacuum Rabi
frequency of the dipole. This condition is much weaker than strong
coupling, and amounts to simply achieving large Purcell factors.
Thus, we can observe transparency in the weak coupling regime.  We
describe how this effect can be useful for designing quantum
repeaters for long distance quantum communication.
\end{abstract}

\pacs{Valid PACS appear here}% PACS, the Physics and Astronomy
\maketitle

%\section{\label{sec:Introduction} Introduction}

The field of cavity quantum electrodynamics (CQED) has seen rapid
progress in the past several years.  One of the main reasons for
this is the development of high quality factors optical
micro-cavities with mode volumes that are less than a cubic
wavelength of light~\cite{VuckovicYamamoto03}.  These high-Q
cavities allow previously unattainable interaction strengths
between a cavity mode and a dipole emitter such as a quantum dot.

There are a large number of applications that require strong
interactions between a cavity and dipole emitter.  These include
methods for conditional phase shifts on single
photons~\cite{DuanKimble04}, single photon
generation~\cite{KuhnHennrich02}, and quantum
networking~\cite{CiracZoller97}. These applications either exploit
modification of the dipole emission rate, or cavity spectrum, when
the two systems are coupled.  It is often perceived that in order
to observe significant modification of the cavity spectrum, one
must enter the so-called ``strong coupling'' regime. In this
regime the interaction strength between the cavity and dipole is
sufficiently large to fully split the cavity mode into a lower and
upper polariton.

In this paper we show that the strong coupling regime is not
required in order to see significant modification of the cavity
spectrum.  We consider a single cavity that is coupled to two
waveguides and behaves as a resonant drop filter. When an optical
field whose frequency is resonant with the cavity is sent down one
waveguide, the drop filter cavity would normally transmit all the
field from one waveguide to another. Hence, the waveguide would
appear opaque at the cavity resonance because all the light would
be dropped to the other port. We show that if one places a
resonant dipole in the drop-filter cavity, the waveguide becomes
highly transparent, even in the weak coupling regime. This
transparency is caused by destructive interference of the two
cavity dressed states.  We refer to this effect as Dipole Induced
Transparency (DIT), because of its close analogy to
Electromagnetically Induced Transparency (EIT) in atomic
media~\cite{HarrisField90}.

The fact that we do not need strong coupling to modify the
transmission of a waveguide is extremely important for the field
of semiconductor CQED.  Although photonic crystal cavities allow
us to approach the regime of strong coupling with a single
emitter, it is very difficult to fabricate cavities that have
sufficiently high quality factors to reach the strong coupling
regime. Things become even more difficult when we attempt to
integrate these cavities with waveguides.  The cavity-waveguide
coupling rate must be sufficiently large that we do not lose too
much of the field to leaky modes.  At the same time, leakage into
the waveguide introduces additional losses making strong coupling
even more difficult to achieve.  Thus strong coupling and
efficient waveguide interaction require mutually conflicting
demands on the performance of the cavity.  Our result relaxes the
constraint on strong coupling, allowing us to work in a practical
parameter regime.  To demonstrate the application of DIT, we
conclude this paper by showing how it can be used to share
entanglement between spatially separated dipoles, and to perform a
full non-destructive Bell measurement on two dipoles. These
operations are extremely useful for building quantum
repeaters~\cite{BriegelDur98,DuanLukin01}.

\begin{figure}
\centering\includegraphics[width=5cm]{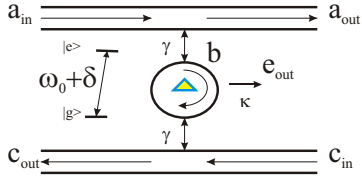} \caption{Cavity
waveguide system for quantum repeaters.} \label{fig:CavWguide}
\end{figure}

Fig.~\ref{fig:CavWguide} shows a schematic of the type of system
we are considering. A cavity containing a single dipole emitter is
evanescently coupled to two waveguides. The cavity is assumed to
have a single relevant mode, which couples only to the forward
propagating fields (e.g. a whispering gallery mode).  This system
is equivalent to an input field reflecting off of a double-sided
linear cavity, and our analysis equally applies to both cases. The
dipole may be detuned by $\delta$ from cavity resonance, denoted
$\omega_0$, while $g$ is the vacuum Rabi frequency of the dipole.
Both waveguides are assumed to have equal coupling rate into the
cavity. This condition is known as critical coupling, and should
result in the input field from one waveguide being completely
transmitted to the other when
$\gamma\gg\kappa$~\cite{ManolatouKhan99}.  When a dipole is placed
inside the cavity, the cavity mode will split into two modes, the
lower and upper polariton branches, that are shifted from the
center frequency by the vacuum Rabi frequency. In the strong
coupling regime, the vacuum Rabi frequency is sufficiently large
that the cavity mode is split by more than a linewidth.  In this
regime, the cavity spectrum is no longer resonant with the input
field, which now remains in its original waveguide.  Our main
interest, however, is in the weak coupling regime where the vacuum
Rabi frequency does not exceed the cavity decay rate. In this
case, the lower and upper polariton branches overlap
significantly, and are still largely resonant with the input
field.  Nevertheless, the two branches can still destructively
interfere in a narrow spectral region near zero detuning.  This
interference is analogous to the interference between the two
dressed states of an atomic lambda system in Electromagnetically
Induced Transparency.

To establish this, we begin with the Heisenberg operator equations
for the cavity field operator $\b$ and dipole operator $\sigm$,
given by~\cite{WallsMilburn}
  \begin{eqnarray}
    \frac{d\b}{dt} & = & -\left(i\omega_0 + \gamma + \kappa/2 \right)\b  -
    \sqrt{\gamma} \left(\ain + \cin\right) \nonumber\\
     & &  - \sqrt{\kappa}\ein -
    ig\sigm \label{eq:Heisenbergb}\\
    \frac{d\sigm}{dt} & = & -\left(i\left(\omega_0+\delta\right)+\frac{\tau}{2}\right)\sigm + ig\sigz\b-\f \label{eq:HeisenbergSig}
  \end{eqnarray}
The operators $\ain$ and $\cin$ are the field operators for the
flux of the two input ports of the waveguide, while $\ein$ is the
operator for potential leaky modes. The bare cavity has a resonant
frequency $\omega_0$ and an energy decay rate $\kappa$ (in the
absence of coupling to the waveguides). This decay rate is related
to the cavity quality factor Q by $\kappa=\omega_0/Q$. The
parameter $\gamma$ is the energy decay rate from the cavity into
each waveguide. Similarly, the dipole operator $\sigm$ has a decay
rate $\tau$, and $\f$ is a noise operator which preserves the
commutation relation. The output fields of the waveguide, $\aout$
and $\cout$, are related to the input fields
by~\cite{WallsMilburn}
  \begin{eqnarray}
    \aout - \ain & = & \sqrt{\gamma}\b \label{eq:ascat}\\
    \cout - \cin & = & \sqrt{\gamma}\b \label{eq:cscat}
  \end{eqnarray}

Eq.~\ref{eq:HeisenbergSig} is difficult to solve because the field
operator $\b$ is multiplied by the time varying operator $\sigz$.
However, we can significantly simplify the problem by looking at
the weak excitation limit, where the quantum dot is predominantly
in the ground state. In this limit, $\langle\sigz(t)\rangle\approx
-1$ for all time, and we can substitute $\sigz(t)$ with its
average value of $-1$. After deriving a solution, we will check
the validity of this approximation.

Assuming the cavity is excited by a weak monochromatic field with
frequency $\omega$, we calculate the response of $\b$ and $\sigm$
using fourier decomposition.  We assume that the cavity decay rate
is much faster than the dipole decay rate, so that
$\tau/\gamma\approx 0$. This is a realistic assumption for a
quantum dot coupled to a photonic crystal cavity, but does not
necessarily apply in atomic systems coupled to very high-Q optical
resonators.  In this limit the waveguide input-output relations
are given by the expressions
    \begin{eqnarray}
    \aout & = & \frac{-\gamma \cin + \left( -i\Delta\omega + \frac{\kappa}{2} +\frac{g^2}{-i\left(\Delta\omega - \delta\right)
    +\tau/2}\right)\ain
    - \sqrt{\kappa\gamma}\ein}{-i\Delta\omega + \gamma + \kappa/2 + \frac{g^2}{-i\left(\Delta\omega - \delta\right)+\tau/2}} \label{eq:asolved}\\
    \cout & = & \frac{-\gamma \aout + \left( -i\Delta\omega + \frac{\kappa}{2} + \frac{g^2}{-i\left(\Delta\omega - \delta\right)+\tau/2}\right)\cout
    -\sqrt{\kappa\gamma}\eout}{-i\Delta\omega + \gamma + \kappa/2 + \frac{g^2}{-i\left(\Delta\omega
    - \delta\right)+\tau/2}} \label{eq:csolved}
  \end{eqnarray}
where $\Delta\omega = \omega - \omega_0$.

First, consider the case where the dipole is resonant with the
cavity, so that $\delta=0$.  In the ideal case, the bare cavity
decay rate $\kappa$ is very small and can be set to zero. In this
limit, when the field is resonant with the cavity and $g=0$ we
have $\ain=-\cout$, as one would expect from critical coupling. In
the opposite regime, when $2g^2/\tau\gg\gamma+\kappa/2$ we have
$\ain=\aout$, so that the field remains in the original waveguide.
This condition can be re-written as $F_p = 2g^2/[(\gamma +
\kappa/2)\tau]\gg1$, where $F_p$ is the Purcell factor.  Thus, in
order to make the waveguide transparent, we need to achieve large
Purcell factors.  However, we do not need the strong coupling
regime $(g>\gamma+\kappa/2)$. When $\tau\ll\gamma+\kappa/2$ we can
achieve transparency for much smaller values of $g$.  In this
sense, our scheme is best suited for implementation in photonic
crystal cavities coupled to quantum dots.  The small mode volumes
of photonic crystal cavities, coupled with the large oscillator
strength of quantum dots, allows us to achieve the large Purcell
factors needed for proper
operation~\cite{EnglundFattal05,BadolatoHennessy05,VuckovicFattal03}.
The above condition has another interpretation that can be
borrowed from atomic physics. The critical atom number
$N_0=(2\gamma +\kappa)\tau/g^2$ and critical photon number
$m_0=(\tau/2g)^2$ are defined as the number of atoms and photons
in the cavity required to see modification of the cavity
spectrum~\cite{Kimble}. Our condition is equivalent to $N_0\ll 1$,
so a single emitter is enough to modify the cavity. Also, because
$\tau\ll g$ we automatically have $m_0\ll 1$.

We now go back and check the validity of our assumption that
$\langle\sigz\rangle\approx -1$, which is equivalent to stating
that $\langle\sigp\sigm\rangle\ll 1$. Using
Equations~\ref{eq:Heisenbergb}-\ref{eq:cscat}, and assuming
$F_p\gg 1$, we can show that on resonance,
$\langle\sigp\sigm\rangle\ll 1$ is equivalent to the condition
$\langle\aindag\ain\rangle\ll g^2/\gamma$. This condition
basically states that the incoming photon flux
$\langle\aindag\ain\rangle$ must be much smaller than the modified
spontaneous emission decay rate of the emitter (in the limit that
the cavity decay is dominated by $\gamma$), and is well satisfied
in the operating regime we are working in.

\begin{figure}
\centering\includegraphics[width=5cm]{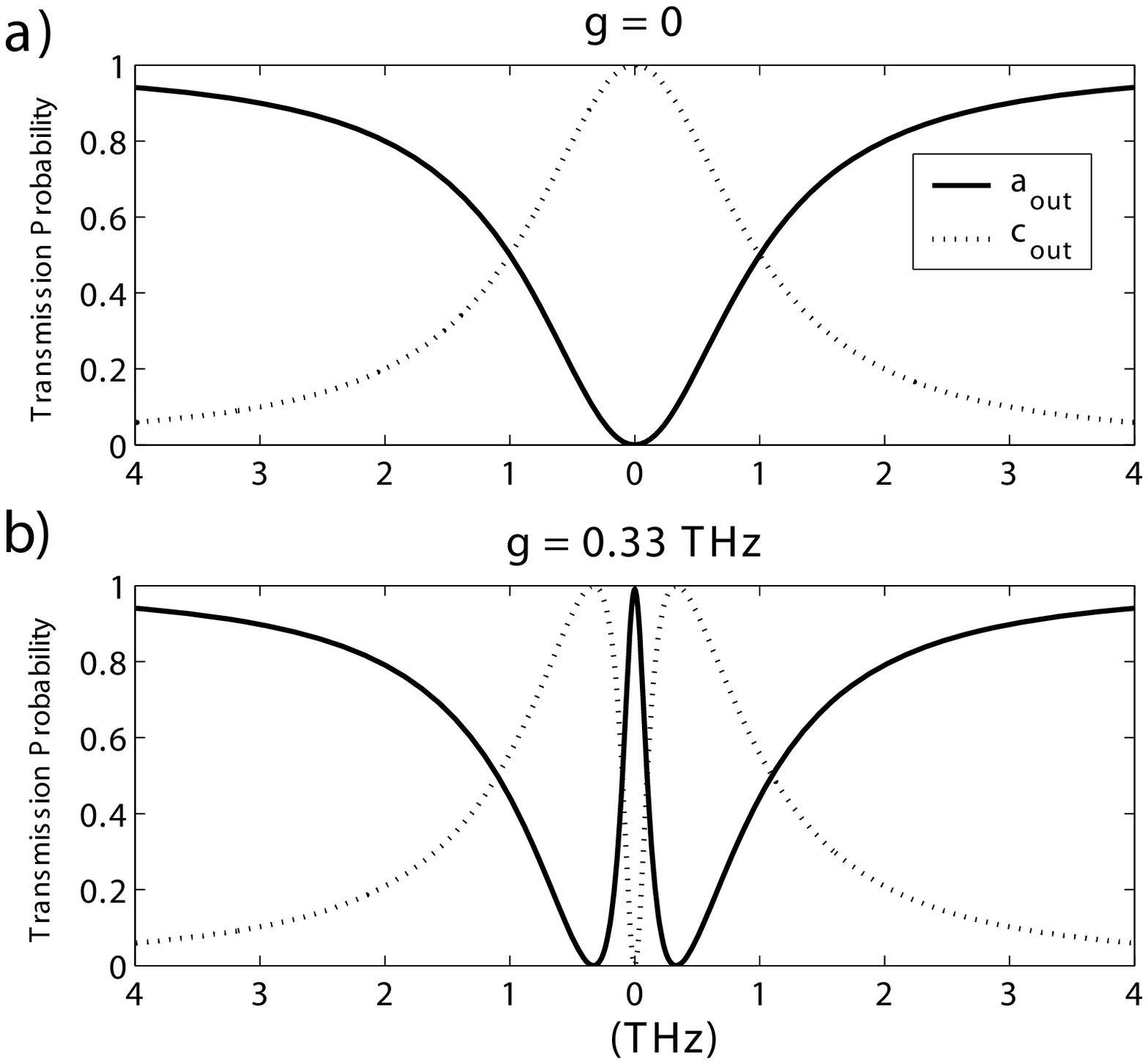}
\caption{Probability for field in $\ain$ to transmit into $\aout$
and $\cout$ respectively. (a) transmission with no dipole in
cavity. (b) transmission with a dipole in the cavity}
\label{fig:scatter}
\end{figure}

Fig.~\ref{fig:scatter} plots the probability that $\ain$ transmits
into $\aout$ and $\cout$.  We use realistic experimental
parameters to create this plot.  We set $\gamma=1THz$ which is
about a factor of 10 faster than $\kappa$ for a cavity with a
quality factor of $Q=10,000$.  We set $g=330GHz$, a number
calculated from FDTD simulations of cavity mode volume for a
single defect dipole cavity in a planar photonic crystal coupled
to a quantum dot~\cite{VuckovicYamamoto03}. The dipole decay rate
is set to $\tau=1GHz$, taken from experimental
measurements~\cite{VuckovicFattal03}.

Panel (a) of Fig.~\ref{fig:scatter} considers the case where the
cavity does not contain a dipole.  In this case $g=0$,
representing a system where two waveguides are coupled by a
cavity.  This well known structure is often referred to as a drop
filter. The width of the transmission spectrum for the drop filter
is determined by the lifetime of the cavity, which in our case is
dominated by $\gamma$.

When a dipole is present in the cavity, the result is plotted in
panel (b).  In this case, a very sharp peak in the transmission
spectrum appears at $\domega=0$.  This peak is caused by
destructive interference of the cavity field, which prevents the
input field from entering the cavity. On resonance, all of the
field is now transmitted through the waveguide instead of being
dropped to the other port. The spectral width of the transmission
peak is roughly equal to $g$. It is important to note that the
transmission is almost complete, even though $g$ is a factor of 3
smaller than the cavity decay rate of $\gamma+\kappa/2$.

\begin{figure}
\centering\includegraphics[width=5cm]{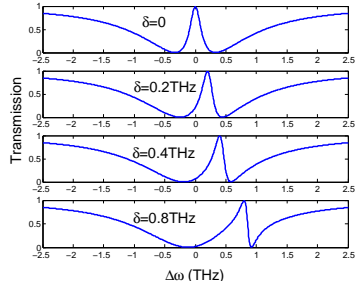}
\caption{Transmission of waveguide as function of $\delta$, the
detuning of the dipole from the cavity.} \label{fig:DeltaPlot}
\end{figure}

We now consider the effect of detuning the dipole.  The
transmission spectrum for several values of $\delta$ is plotted in
Fig~\ref{fig:DeltaPlot}.  Introducing a detuning in the dipole
causes a shift in the location of the transmission peak., so that
destructive interference occurs when the field frequency is equal
to the dipole frequency.  Thus, we do not have to hit the cavity
resonance very accurately to observe DIT.  We only need to overlap
the dipole resonance within the cavity transmission spectrum.

\begin{figure}
\centering\includegraphics[width=5cm]{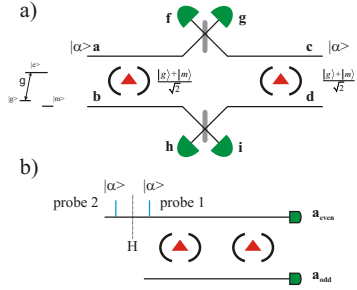}
\caption{Application of DIT to quantum repeaters. a) a method for
generating entanglement between two dipoles using DIT.  b) a
non-destructive Bell measurement.} \label{fig:RepeaterFig}
\end{figure}

The fact that we can strongly modify the transmission spectrum of
a waveguide by the state of a dipole can be extremely useful for
quantum information processing.  As one example, we now present a
way in which DIT can be applied to engineering quantum repeaters
for long distance quantum communication.  Quantum repeaters can be
implemented all optically~\cite{PanGasparoni03,WaksZeevi02}, as
well as using atomic systems~\cite{DuanLukin01}. One of the main
problems with these proposals is that it is difficult to implement
the full Bell measurement required for swapping entanglement.
This leads to a communication rate that is exponentially decaying
with the number of repeaters. More recent proposals incorporate
interaction between nuclear and electron spins to implement the
full Bell measurement~\cite{ChildressTaylor05}.  Here we propose a
method for implementing entanglement, as well as a full Bell
measurement on an atomic system using only interaction with a
coherent field. This leads to an extremely simple implementation
of a quantum repeater.

\begin{figure}
\centering\includegraphics[width=5cm]{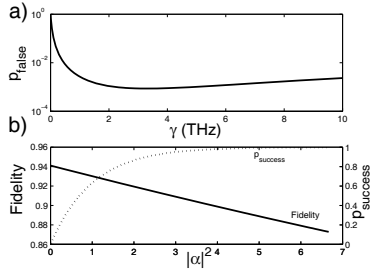} \caption{Panel
(a), probability of detecting even parity for an odd parity state
as a function of $\gamma$. Panel (b), solid line plots the
fidelity of the state $(\|gg>\pm\|mm>)/{\sqrt{2}}$ after a parity
measurement. Dotted line plots the probability that the measuring
field contains at least one photon for detection.
\label{fig:FidSuc}}
\end{figure}

In panel (a) of Fig.~\ref{fig:RepeaterFig}  we show how DIT can be
used to generate entanglement between two spatially separated
dipoles.   A weak coherent beam is split on a beamsplitter, and
each port of the beamsplitter is then sent to two independent
cavities containing dipoles. The waveguide fields are then mixed
on a beamsplitter such that constructive interference is observed
in ports $\f$ and $\h$.  Each dipole is assumed to have three
relevant states, a ground state, an excited state, and a long
lived metastable state which we refer to as $\|g>$, $\|e>$, and
$\|m>$ respectively.  The transition from ground to excited state
is assumed to be resonant with the cavity while the metastable to
excited state transition is well off resonance from the cavity,
and is thus assumed not to couple to state $\|e>$.  The states
$\|g>$ and $\|m>$ represent the two qubit states of the dipole.

When the dipole is in state $\|m>$, it does not couple to the
cavity, which now behaves as a drop filter. Thus, we have a system
that transforms $\aindag\|g>\|0>\to\aoutdag\|g>\|0>$ and
$\aindag\|m>\|0>\to-\coutdag\|m>\|0>$.  This operation can be
interpreted as a C-NOT gate between the state of the dipole and
the incoming light.  When the dipole is in a superposition of the
two states, this interaction generates entanglement between the
path of the field and the dipole state.  After the beamsplitter,
this entanglement will be transferred to the two dipoles. If the
state of both dipoles is initialized to $(\|g> + \|m>)/\sqrt{2}$,
it is straightforward to show that a detection event in ports $\g$
or $\i$ collapses the system to $(\|g,m> - \|m,g> )/\sqrt{2}$.

Another important operation for designing repeaters is a Bell
measurement.  Panel (b) of Fig.~\ref{fig:RepeaterFig} shows how to
implement a complete Bell measurement between two dipoles using
only cavity waveguide interactions with coherent fields.  The two
cavities containing the dipoles are coupled to two waveguides.
When a coherent field $\|\alpha>$ is sent down waveguide 1, each
dipole will flip the field to the other waveguide if it is in
state $\|m>$, and will keep the field in the same waveguide if it
is in state $\|g>$. Thus, a detection event at ports $\aeven$ and
$\aodd$ corresponds to a parity measurement.  A Bell measurement
can be made by simply performing a parity measurement on the two
dipoles, then a Hadamard rotation on both dipoles, followed by a
second parity measurement.

To understand why this works, consider the four Bell states
$\|\phi_\pm> = (\|gg>\pm\|mm>)/\sqrt{2}$ and $\|\psi_\pm> =
(\|gm>\pm\|mg>)/\sqrt{2}$.  The first parity measurement
distinguishes the states $\|\phi_\pm>$ from $\|\psi_\pm>$, since
these two groups have opposite parity.  After a Hadamard rotation
on both dipoles, it is easy to verify that the states $\|\phi_+>$
and $\|\psi_->$ are unaffected, while $\|\phi_->\to\|\psi_+>$ and
$\|\psi_+>\to\|\phi_->$, and thus flip parities. The second
measurement will then distinguish between the states $\|\phi_+>$
from $\|\phi_->$ and $\|\psi_+>$ from $\|\psi_->$, which
completely distinguish the four Bell states. It is important to
note that this measurement is non-destructive, in that after the
measurement the state of the dipoles remains in the measured
state.

The performance of the Bell apparatus is analyzed in
Fig~\ref{fig:FidSuc}.  Panel (a) plots the probability that an odd
parity state will falsely create a detection event in port
$\aeven$, as a function of $\gamma$.  The probability becomes high
at large $\gamma$ due to imperfect transparency.  It also
increases at small $\gamma$ because of imperfect drop filtering.
The minimum value of about $10^{-3}$ is achieved at approximately
3THz.  In panel (b) of Fig.~\ref{fig:FidSuc} we plot both the
fidelity and success probability of a parity measurement as a
function of the number of photons in the probe field.  The
fidelity is calculated by applying the Bell measurement to the
initial state $\|\psi_i>=(\|g,g> \pm \|m,m>)/\sqrt{2}$, and
defining the fidelity of the measurement as
$F=|\bk<\psi_f|\psi_i>|^2$, where $\|\psi_f>$ is the final state
of the total system which includes the external reservoirs.  The
probability of success is defined as the probability that at least
one photon is contained in the field.  The fidelity is ultimately
limited by cavity leakage, which results in ``which path''
information beaing leaked to the environment. This information
leakage depends the strength of the measurement which is
determined by the number of photons in the probe fields. Using
more probe photons results in a higher success probability, but a
lower fidelity.  To calculate this tradeoff, we use previously
described values for cavity and reservoir losses, and set the
coupling rate $\gamma$ to 4THz, which is where the probability of
false detection is near its minimum.   At an average of three
photons, a fidelity of over $90\%$ can be achieved with a success
probability exceeding $95\%$.  These numbers are already
promising, and improved cavity and dipole lifetimes could lead to
even better operation.

This work was funded in part by the MURI center for photonic
quantum information systems (ARO/DTO Program DAAD19-03-1-0199),
and a Department of Central Intelligence postdoctoral grant.

\end{document}